\documentclass[12pt]{scrartcl}
\usepackage{amsmath}
\usepackage{amssymb}
\usepackage{amsthm}
\usepackage{slashed}
\usepackage{graphicx}
\usepackage{color}
\usepackage[centering,headsep=15pt,head=80pt,foot=30pt,margin=70pt]{geometry}
\usepackage[scaled]{helvet}
\usepackage[slantedGreek]{mathptmx}

\newcommand{\Group}[2]{{ \hbox{{\itshape{#1}}($#2$)} }}
\newcommand{\U}[1]{\Group{U\kern0.05em}{#1}}
\newcommand{\SU}[1]{\Group{SU\kern0.1em}{#1}}
\newcommand{\SL}[1]{\Group{SL\kern0.05em}{#1}}
\newcommand{\Sp}[1]{\Group{Sp\kern0.05em}{#1}}
\newcommand{\SO}[1]{\Group{SO\kern0.1em}{#1}}

\newcommand{\mybar}[1]%
    {{\kern 0.8pt\overline{\kern -0.8pt#1\kern -0.8pt}\kern 0.8pt}}
\newcommand{\roughly}[1]%
    {{ \mathrel{\raise.3ex\hbox{ $#1$\kern-.75em\lower1ex\hbox{$\sim$}} } }}

\newcommand{\nop}[1]{:\kern-.3em#1\kern-.3em:}

\newcommand{\al}{\ensuremath{\alpha}}
\newcommand{\be}{\ensuremath{\beta}}
\newcommand{\ga}{\ensuremath{\gamma}}
\newcommand{\Ga}{\ensuremath{\Gamma}}
\newcommand{\de}{\ensuremath{\delta}}

\renewcommand{\th}{\ensuremath{\theta}}

\newcommand{\la}{\ensuremath{\lambda}}

\newcommand{\si}{\ensuremath{\sigma}}

\newcommand{\om}{\ensuremath{\omega}}

\newcommand{\mcl}[1]{\mathcal{#1}}

\numberwithin{equation}{section}
\numberwithin{figure}{section}

\begin{document}
\begin{titlepage}
\begin{flushright}
KUNS-2955
\end{flushright}
\vskip 5em
\begin{center}
{\Large \bfseries
 Stellar cooling limits on light scalar boson revisited\\
}
\vskip 4em
\renewcommand{\thefootnote}{\alph{footnote}}

Yasuhiro Yamamoto$^\sharp$
and
Koichi Yoshioka$^\flat$

\vskip 4em
$^\sharp$
	\textit{Physics Division, National Center for Theoretical Sciences, National Taiwan University, Taipei 10617, Taiwan}\\
$^\flat$
	\textit{Department of Physics, Kyoto University, Kyoto 606-8502, Japan}
\vskip 4em
\textbf{Abstract}
\end{center}
\medskip
\noindent

We revisit the stellar cooling limits on the light scalar boson whose coupling to the Standard Model particles is described by its mixing with the Higgs boson.
Strong constraints have been obtained from the electron-nucleus bremsstrahlung process and the resonant plasma effect in the medium.
We find that the bremsstrahlung contribution from the electron and nucleus scattering is of similar magnitude to the plasma mixing effect including the off-resonant mixing. 
The constraints on the scalar coupling are found to be about three orders of magnitude weaker than the previous evaluations.
For white dwarfs, the stellar cooling constraint is even more suppressed due to the Pauli blocking effect.
We obtain limits on the Higgs-scalar mixing angle of $10^{-10}$--$10^{-9}$ in the region where the scalar mass is lighter than about 10 keV.

\bigskip
\vfill
\end{titlepage}
\tableofcontents
\setcounter{footnote}{0}
\section{Introduction}

A light new scalar boson is a candidate introduced in models beyond the Standard Model.
One of the features of such a scalar boson is that it is always coupled to the Higgs boson through the renormalizable interaction, the so-called Higgs portal coupling.
With this coupling, the interaction between the light scalar boson and the Standard Model is simply described by the mixing angle with the Higgs boson.
Because of this generality to describe a large class of models and the simplicity of being characterized by two parameters, the mixing and the mass, the Higgs portal light scalar has attracted a lot of interest.

Despite the simplicity of the description, the existence of a light new scalar boson could have a variety of experimental and observational signals.
The unknown light boson is a main target of low-energy experiments~\cite{1901.09966}.
Because such a boson can change the low-energy experimental signals, they are often used to explain whack-a-mole anomalies.

The light boson could also affect the evolution of the universe and stars.
One important role is as a mediator in the dark matter sector.
While the interaction with the Standard Model should be small enough to escape experiments and observations, the interaction with the dark matter is poorly constrained.
Then the light boson mediates a large self-interaction of the dark matter and could explain the small-scale structure puzzles~\cite{1508.03339,1705.02358}.
As we study in this paper, another important astrophysical phenomenon is the stellar cooling, namely, the loss of energy in stars through the emission of light bosons.
Various observations of stellar cooling can provide valuable information about the possible presence of new particles and phenomena that could affect their evolution~\cite{Raffelt,1708.02111}.

The search for a light new scalar boson is therefore not only a key focus of many particle physics experiments but also a topic of interest in the fields of cosmology and astrophysics~\cite{1812.07585}.
Using a variety of experimental and observational techniques to try to detect or constrain such a particle, new potential implications of its existence have been actively explored.

The stellar cooling argument is known to provide the strongest constraints on the light new degree of freedom in the mass range from $O(1)$ eV to $O(100)$ keV.
Among various energy emission processes, the resonant plasma mixing is a main processes to constrain the light boson property~\cite{1302.3884,1305.2920}.
For a scalar boson with the Higgs mixing, the resonant plasma production excludes the mixing up to $O(10^{-9})$ for the scalar mass lighter than typical star temperatures, $O$(1--10) keV~\cite{1611.05852}.
Recently, it has been claimed that the scalar bremsstrahlung emission from nuclei in the electron-nucleus ($e$-$N$) scattering provides a dominant contribution even in the region where the scalar boson mass is lower than the plasma frequency~\cite{2010.01124,2205.01669}.

In this paper, we re-evaluate the bremsstrahlung contributions to the stellar cooling limits on the Higgs portal light scalar boson.
We also include the plasma mixing contributions by electron.
The scalar emission rate by electron is found to be consistent with the result of Ref.~\cite{1611.05852}, and the emissions by nuclei discussed in Refs.~\cite{2010.01124,2205.01669} are modified to match to our evaluation.
As a result, the constraints on the mixing angle between the Higgs and the light scalar become weaker by several orders of magnitude.

The rest of this paper is organized as follows.
In the next section, we present the general formulas for the emission and absorption rates of the light scalar by the stellar medium, in particular the $e$-$N$ bremsstrahlung contribution.
The detailed analysis is discussed separately for each celestial object: the Sun, horizontal-branch stars, red giants, and white dwarfs in Section.~\ref{SecCooling}.
Finally, we summarize our results in Section~\ref{SecCon}.

\section{Bremsstrahlung emission of light scalar}
\label{SecBremss}

We here briefly discuss the general formulas for the stellar cooling.
The energy emission $Q$ carried by a light new particle per unit time and volume is described by
\begin{align}
 Q(T) =& \int dE_S E_S \frac{d R}{d E_S} P_\text{op},
\label{Qgeneral}
\end{align}
where $T$ is the stellar temperature and $E_S$ is the energy of the emitted particle, $R$ is the production rate of the light particle and the opacity factor $P_\text{op}$ is given by the absorption rate $\Ga_\text{abs}$ and the decay width $\Ga_\text{dec}$ of the particle.
With these two parameters, the opacity factor is defined as
\begin{align}
 P_\text{op} =& \exp[-d/\la], \\
 \la^{-1} =& \frac{1}{\be} \left( \Ga_\text{abs} +\frac{\Ga_\text{dec}}{\ga} \right),
\end{align}
where $d$ is the propagation distance of the particle and $\be$ and $\ga$ are the velocity and the Lorentz factor of the emitted light particle.
The explicit form of the absorption rate is discussed later in this section and Appendix~\ref{AppBremss}.
The emission luminosity $L$ of the light particle is obtained by integrating over the entire volume of a star,
\begin{align}
 L =& \int d^3r\, Q(T(r), \{n(r)_I\}),
\end{align}
where $\{n(r)_I\}$ is the number densities of particles contributing to the emission process in the star.

The light scalar boson production rate $R$ of the $n_I\to n_F +1$ scattering is 
\begin{align}
 R =& \int d\Pi_{n_I+n_F+1} (2\pi)^4 \de^4(\{p_I\} -k_S -\{p_F\}) \Bigl(\prod_{I,F} n_I (1-f_F) f_I\Bigr) \overline{\sum_\text{spin}}  |\mcl{M}|^2 ,
 \label{Rgeneral}.
\end{align}
where $k_S$ is the scalar momentum, $\mcl{M}$ is the amplitude of the production process and $n_{I(F)}$, $\{p_{I(F)}\}$ and $f_{I(F)}$ denote the number densities, the sum of the momenta and the distribution functions of the initial (final) state, respectively.
Following Refs.~\cite{RefAbs,hep-ph/0505090}, we write the absorption rate with the amplitude $\mcl{M}'$ where the light scalar is flipped to the initial state,
\begin{align}
  \Ga_\text{abs} = \frac{1}{2E_S} \int d\Pi_{n_I+n_F} (2\pi)^4 \de^4(\{p_I\}+k_S -\{p_F\}) \Bigl(\prod_{I,F} n_I (1-f_F) f_I\Bigr) \overline{\sum_\text{spin}}  |\mcl{M}'|^2 .
\end{align}
If there are no particles other than the light scalar in the initial state, this formula is identical to the decay width in the scalar's rest frame.

\paragraph{Energy emission by $e$-$N$ bremsstrahlung}

We use the following effective Lagrangian for the light scalar boson $S$ with its mass $m_S$ and the Standard Model fermions, electron $e$ and nucleons $p,n$,
\begin{align}
 \mcl{L} = -g_e S \bar{e}e -g_p S \bar{p}p -g_n S \bar{n}n.
\end{align}
In the Higgs portal light scalar model, these couplings are given by their Yukawa couplings to the Higgs boson and the mixing angle between the Higgs and light scalar boson as $g_{e,p,n}=(y_{e,p,n}/\sqrt{2}) \sin\th$.
The coupling between $S$ and two photons is small enough to be neglected in both the production rate and the opacity factor in this model.
As for heavy nucleus, we assume that its constituent nucleons can be treated coherently, since typical stellar temperatures are well below the binding energy of nucleus.

The energy emission rate by the $e$-$N$ bremsstrahlung process can be analytically simplified if electrons are non-relativistic and non-degenerate.
In this case, we can use the Maxwell-Boltzmann distribution for the initial state energy distribution.
Without the opacity factor, the energy emission rate is given by
\begin{align}
  Q^{eN} 
 =&
  \frac{2^3 Z^2 \al^2 n_e n_N T^{1/2}}{3(2\pi)^{3/2} m_N^2 m_e^{3/2}} (g_N m_e -g_e m_N)^2\  I(m_S/T) ,
\label{EqBremss}
\end{align}
where $n_{e,N}$ are the number densities of electron and nucleus, $m_{e,N}$ are their masses, $\al$ is the fine structure constant and $Z$ is the atomic number of the nucleus $N$.
The coupling between the light scalar and the nucleus is given by $g_N=Z g_p+(A-Z) g_n$ with the mass number $A$.
The integration $I(y)$ is defined as
\begin{align}
 I(y) =\int^\infty_y du \int^\infty_0 dv \int^1_{-1} dz \int^\infty_y dx\  e^{-u}
       \sqrt{uv} \left(1-\frac{y^2}{x^2}\right)^{3/2} \frac{\de(u-v-x)}{u+v-2\sqrt{uv}z},
\label{EqIntbremss}
\end{align}
with $x=E_S/T$.
For other variables, we have assigned the momenta of the initial and final state particles as $N(p_1) e(p_2) \to N(p'_1) e(p'_2) S(k_S)$ and defined the relative momenta as $\vec{p}_{I(F)}=(m_e \vec{p}_1^{(\prime)} -m_N \vec{p}_2^{(\prime)})/(m_e+m_N)$.
Then, the integration variables are $u,v=\vec{p}_{I,F}{}^2/(2m_e T)$ and $z$ is the angle between $\vec{p}_I$ and $\vec{p}_F$~\footnotemark.
If the scalar mass is neglected, the integration Eq.~\eqref{EqIntbremss} becomes $I(0)=2$.
\footnotetext{
Note that our definitions of $u,v$ are slightly different from those used in Refs.~\cite{2010.01124,2205.01669}, while they have a similar integration for the emission rate.
We use $m_e$ instead of $m_N$ for the masses of the relative momenta.
This is introduced as the leading order of the reduced mass of the initial and final states.
}

In the Higgs portal model, the nuclear contribution is slightly smaller than that of electron since the nucleon is not an elementary particle, i.e. $(y_e/y_{p,n})(m_{p,n}/m_e)\sim 9/2$~\cite{RefSvz}.
Assuming that the coupling and the mass of nucleus are naively scaled by the mass number $A$ in this model, the emission rate is independent of the nuclear species except for the overall scaling $Z^2$, as seen from Eq.~\eqref{EqBremss}.
In other words, succeeding to the non-decoupling effect of the Higgs boson, a heavier nucleus induces a larger emission rate in the Higgs portal scalar model.

As for the part of the bremsstrahlung emission by nucleus in Eq.~\eqref{EqBremss}, the multiplication of the factor $(m_N/m_e)^2$ seems to lead to an expression similar to the previous results shown in Refs.~\cite{2010.01124,2205.01669}.
This difference gives us weaker constraints on the scalar coupling, which is simply suppressed by the electron-nucleus mass ratio.
Without this suppression, the nuclei would have to emit a light scalar boson by the bremsstrahlung process even if $m_N\to\infty$.
Note that, in this heavy mass limit, the static electric field of a nucleus induces the bremsstrahlung process by electron as indicated by Eq.~\eqref{EqBremss}.

\paragraph{Absorption by inverse $e$-$N$ bremsstrahlung}

The absorption rate by the $e$-$N$ bremsstrahlung process is calculated with the diagrams where the light boson line is flipped from the outgoing to the incoming.
As assumed in the emission rate, if the Standard Model fermions are non-relativistic and non-degenerate, we can obtain the following simple formula of the absorption rate,
\begin{align}
   \Ga_\text{abs}
 = \frac{2^2 (2\pi)^{1/2}Z^2 \al^2 n_N n_e}{3m_N^2 m_e^{3/2}T^{7/2}}(g_N m_e -g_e m_N)^2 H(x,m_S/T),
\end{align}
where the function $H(x,y)$ is defined as
\begin{align}
 H(x,y) =\frac{x^2-y^2}{x^5}
 \int_0^\infty du \int_y^\infty dv \int_{-1}^1 dz\  e^{-u}\sqrt{uv} \frac{\de(u-v+x)}{u+v-2\sqrt{uv}z}.
\end{align}
Here the definition of the variables is the same as for the emission rate.
This absorption rate also has the $(m_e/m_N)^2$ suppression for the nuclear interaction part.
In the Higgs portal model, similar to the emission rate, the absorption by electron is slightly larger than the nuclear one and roughly proportional to $Z^2$ for heavy nuclei.

\section{Stellar cooling limits}
\label{SecCooling}

We discuss the stellar cooling limits to the mass and mixing angle of the Higgs portal light scalar boson by the Sun, red giants, horizontal-branch stars and white dwarfs.
To obtain the limits, we assume that these stars, other than the Sun, can be described by the constant temperatures and densities near their cores shown in Table~\ref{TabStar}.
For the Sun, we evaluate the limit with the temperature and the density profiles given by the standard solar model of Refs.~\cite{RefStandardsun}, which are employed in Ref.~\cite{2205.01669}.
\begin{table}[htb]
\caption{
 The typical values for the core compositions, the temperatures $T$, the election number densities $n_e$, the radii $R$, and the luminosity limits $L$ for the Sun, horizontal-branch stars, red giants and white dwarfs.
 The luminosity limits are normalized by the solar luminosity $L_\text{Sun}=4\times 10^{33}$ erg/sec.
For the details of these values, see Ref.~\cite{2010.01124} and references therein, except for the core temperature of white dwarfs.
}
\label{TabStar}
\centering
\begin{tabular}{l|ccccc}
\hline
  Star & Composition & $T$ (keV) & $n_e$ (cm${}^{-3}$) & $R$ (cm) & $L/L_\text{Sun}$ \\
\hline\hline
  Sun & H: 75\%, ${}^4$He: 25\% & 1 & $10^{26}$ & 7$\times 10^{10}$ & $0.03$ \\
	HBs & ${}^4$He & 8.6 & $3\times 10^{27}$ & $3.6\times 10^9$ & $5$ \\
  RGs & ${}^4$He & 10 & $3\times 10^{29}$ & $6\times 10^8$ & $2.8$ \\
  WDs & ${}^{12}$C: 50\%, ${}^{16}$O: 50\% & 1--6 & $10^{30}$ & $10^9$ & $10^{-5}-0.03$ \\ \hline
\end{tabular}
\end{table}

In addition to the $e$-$N$ bremsstrahlung contribution, we include the plasma mixing effect by electrons in the medium.
Following Refs.~\cite{1302.3884,1305.2920,1611.05852}, for the non-relativistic and non-degenerate plasma, this effect is described by the damping rate of $S$,
\begin{align}
  \Ga_S = \frac{g_e^2}{4\pi\al} \frac{\vec{k}_S^2 E_S^2 \si_L}{(E_S \si_L)^2 +(E_S^2 -\om_p^2)^2},
\end{align}
where $\om_p$ is the electron plasma frequency.
The dumping rate of the longitudinal mode of photon $\si_L$ is calculated by the Thomson scattering and the inverse bremsstrahlung process in the medium~\cite{1305.2920}.
This plasma mixing effect induces both the resonant production at $E_S\sim\om_p$ and the off-resonant production.
As discussed below, the resonant production dominates the constraints for the scalar mass lighter than the plasma frequency.
The off-resonant production by the electron plasma gives significant constraints in the region of $m_S>\om_p$.

We include the Debye screening effect in the following calculations, but the limits are only slightly affected by this effect.
To evaluate the absorption effect, one need to integrate over whole region of star taking care of the geometric effect~\cite{2205.01669,2204.11862}.
For simplicity of the calculation, we did not include this effect, because the absorption rate is also corrected by the factor of $(m_e/m_N)^2$.
The effect is estimated as it produced at the center of the stars.
As we see below, the absorption effect does not make gap with the terrestrial bound, while the treatment enhances the effect.

\subsection{Sun}
\label{SecSun}

We discuss the solar production of the Higgs portal light scalar.
Inside the Sun, the plasma is non-relativistic and non-degenerate.
Then, we can use the formulas shown in the previous section.
\begin{figure}[ht]
 \centering
 \includegraphics[scale=.3]{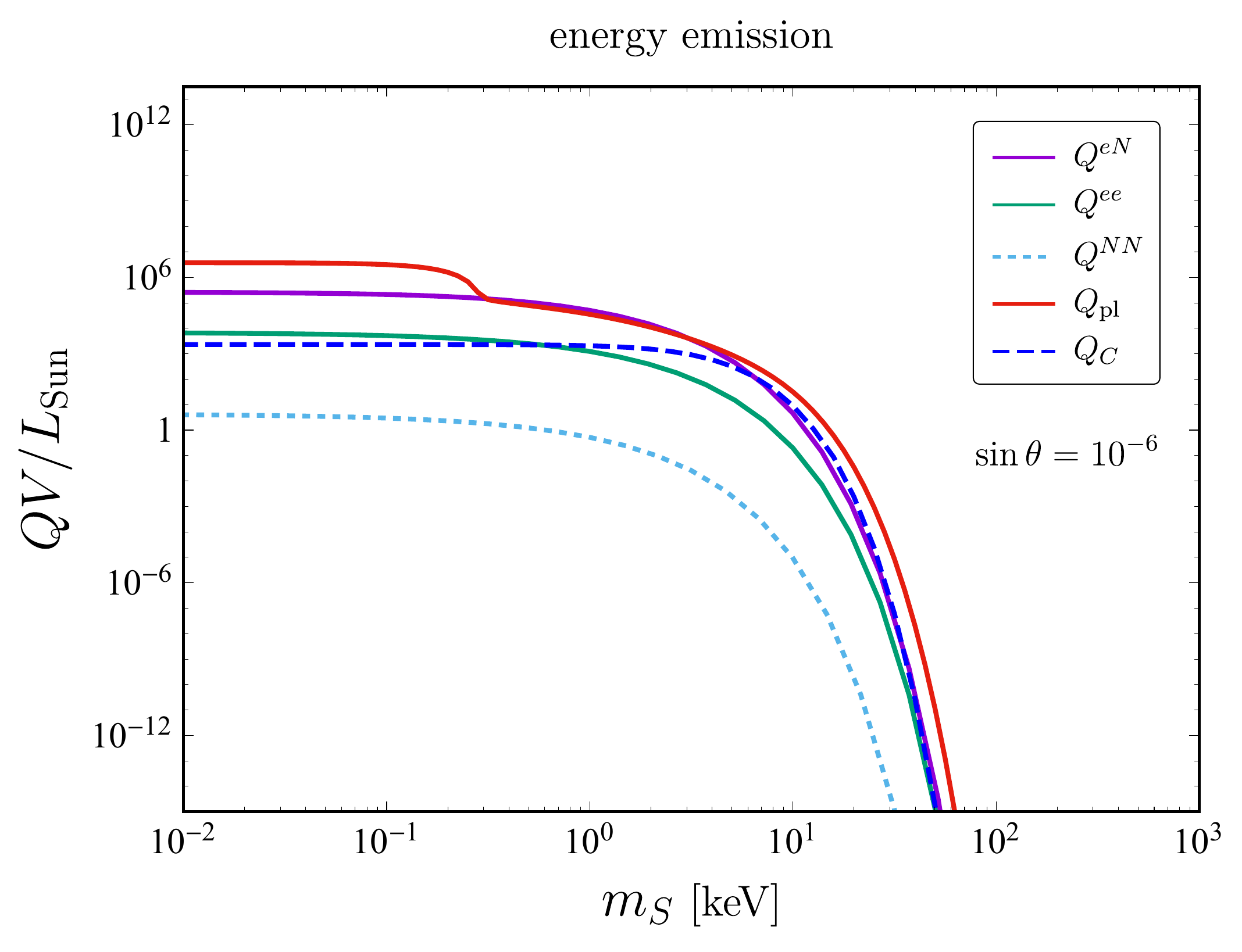}
\caption{
 The scalar energy emission rates normalized by the solar luminosity as the function of the scalar mass in the Sun.
 Each line corresponds to the $e$-$N$ ($Q^{eN}$: purple solid), $e$-$e$ ($Q^{ee}$: green solid), $N$-$N$ ($Q^{NN}$: blue dotted) bremsstrahlung processes, the plasma mixing effect in the medium ($Q_\text{pl}$: red solid), and the Compton-like process $\gamma+e\to S+e\,$ ($Q_C$: blue dashed).
 The mixing angle is fixed as $\sin\th=10^{-6}$ and all lines are scaled as $\sin^2\theta$.
 In the vertical axis, $V$ stands for the solar volume.
} \medskip
\label{FigEmisun}
\end{figure}

In the constant temperature and density approximation shown in Table~\ref{TabStar}, we evaluate the energy emission rate of several processes: the $e$-$N$, $e$-$e$ and $N$-$N$ bremsstrahlung processes, the plasma mixing and the Compton productions in Fig.~\ref{FigEmisun}.
We use the result of the Compton production shown in Ref.~\cite{2010.01124}.
In this figure, we neglect the absorption effect.

The $e$-$N$ bremsstrahlung contribution to the emission rate is about six orders of magnitude smaller than in the previous results. 
Then, in the mass region below the plasma frequency $\sim$0.3 keV, the resonant plasma mixing is found to dominate the energy emission.
The $e$-$N$ bremsstrahlung is about one order of magnitude smaller than the resonant plasma mixing.
In the mass region around 0.3--4 keV, the $e$-$N$ bremsstrahlung process and the off-resonant plasma mixing induce the largest emission rate.
The contributions from the $e$-$e$ bremsstrahlung and the Compton scattering processes are always subdominant.
Due to the nuclear mass suppression, the $N$-$N$ bremsstrahlung process is negligibly small.

\begin{figure}[hbt]
 \centering
 \includegraphics[scale=.3]{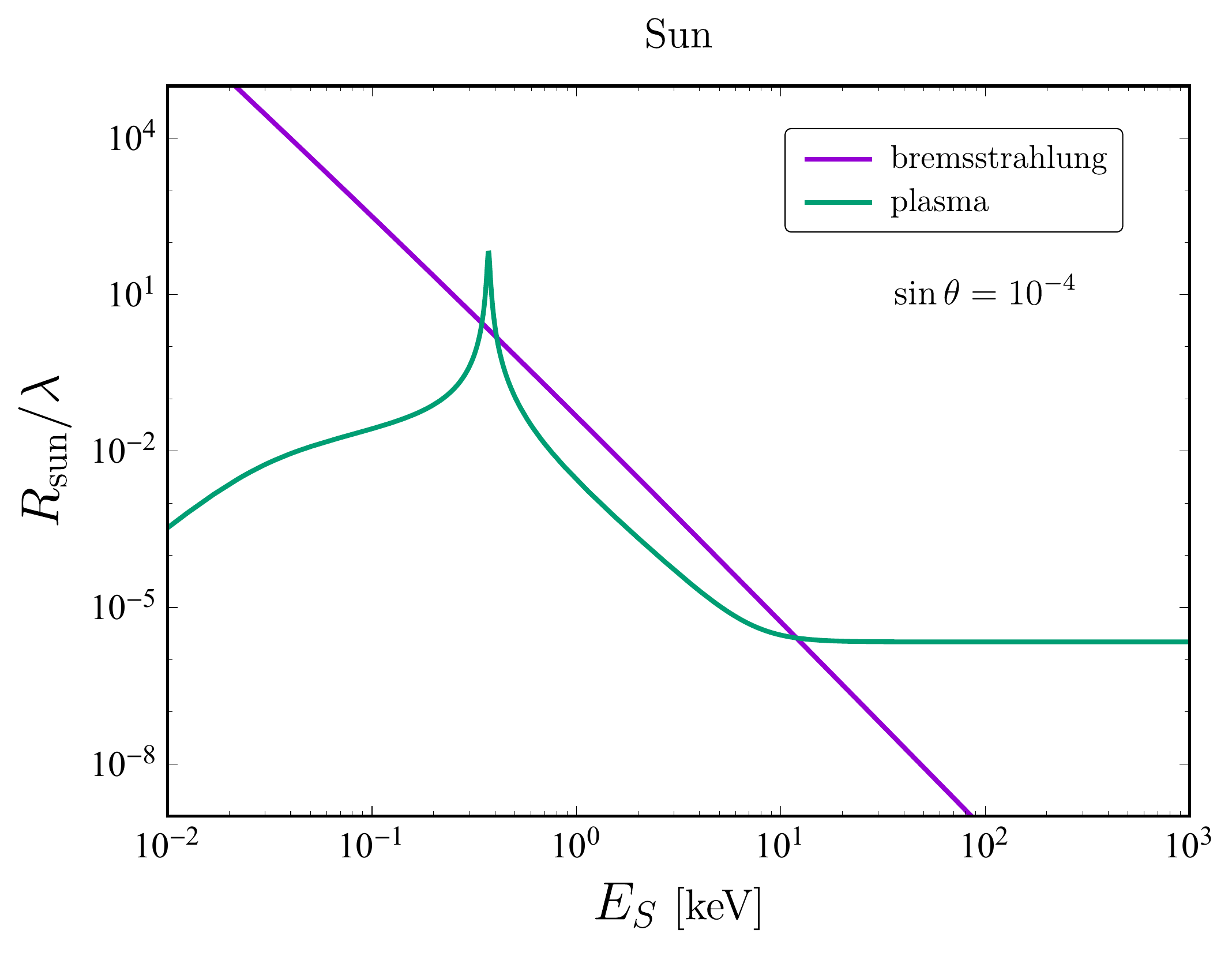}
\caption{
 The ratio of the solar radius to the mean free path of $S$ as the function of the scalar boson energy in the Sun.
 The purple line means the absorption by the $e$-$N$ bremsstrahlung process and the green line is that due to the plasma mixing.
 The plasma mixing effect peaks at the plasma frequency $\simeq 0.29$~keV in the solar environment.
 The mixing angle is fixed as $\sin\th=10^{-4}$ and all lines are scaled as $\sin^2\theta$.
} \medskip
\label{FigMfpsun}
\end{figure}

Fig.~\ref{FigMfpsun} shows the ratio of the solar radius $R_\text{Sun}$ to the mean free path of $S$ in the solar medium as the function of the emitted scalar energy.
In this figure, we set the scalar boson mass $m_S$ small.
For a large $m_S$, there is an energy threshold around $m_S$ below which the opacity factor has no effect.

The plasma mixing contribution has the peak at the plasma frequency where the scalar boson is resonantly absorbed by the plasma.
In the low energy region below this peak, the inverse bremsstrahlung is the dominant process of the absorption.
In the energy region around 1--10 keV, the absorption rates of both processes decrease similarly.
The off-resonant mixing contribution keeps the absorption rate of $\Ga_\text{abs}\sim(g_e^2/4\pi\al) \si_L$ at a higher energy region while the bremsstrahlung contribution almost disappears.
However, such a high-energy particle is not well produced in the Sun.

\begin{figure}[hbt]
 \centering
 \includegraphics[scale=.3]{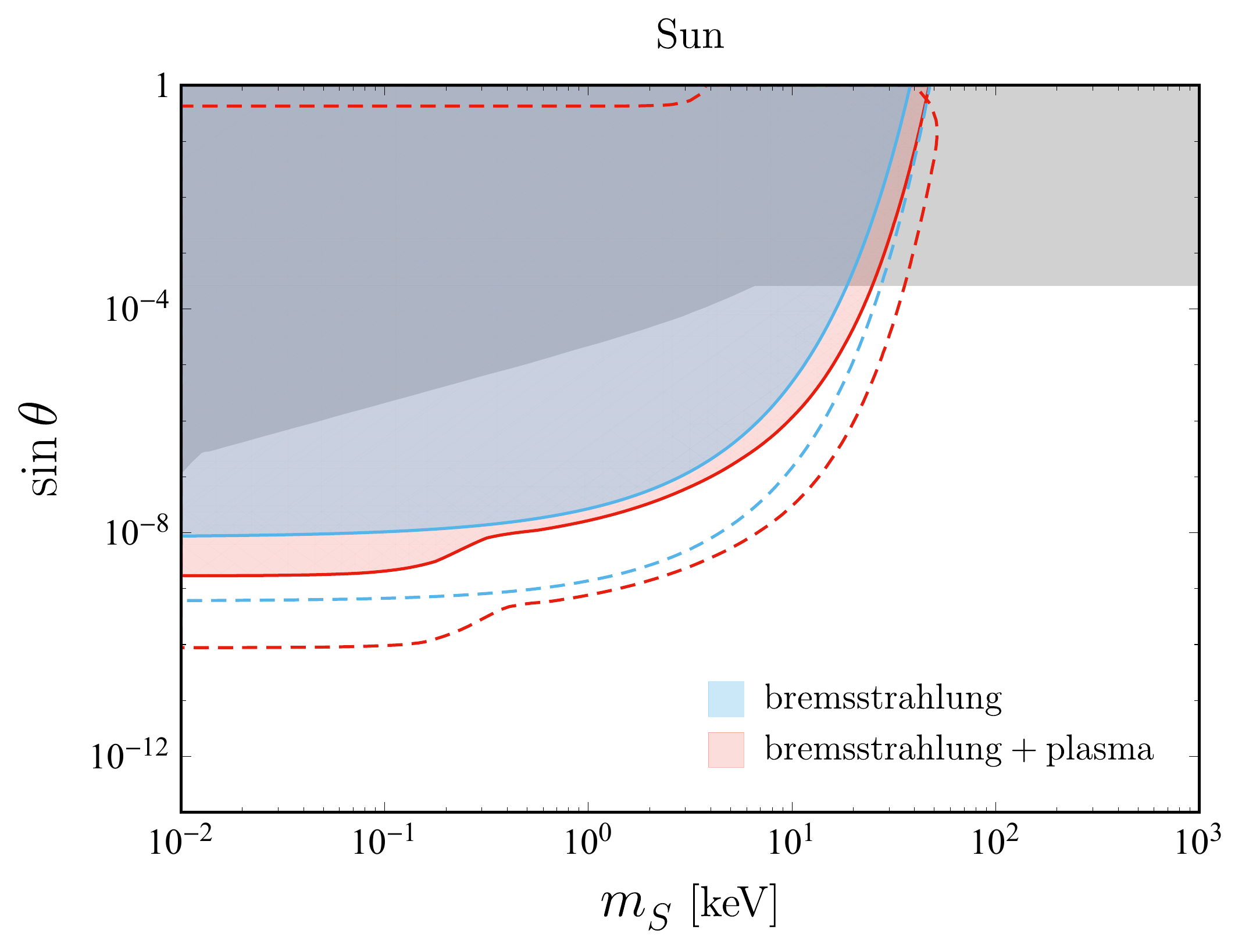}
\caption{
 The limits on the scalar mass $m_S$ and the Higgs-mixing parameter $\sin\theta$ from the Sun.
 The blue region is excluded by the $e$-$N$ bremsstrahlung process.
 The red region is the excluded region including the plasma mixing effect in the stellar medium. 
 The terrestrial experimental bound (gray region) is provided by the meson missing decay~\cite{1809.01876} and the neutron scattering~\cite{2103.05428}.
 The dotted lines are evaluated with the constant solar model of Table~\ref{TabStar}.
} \medskip
\label{FigSun}
\end{figure}

We show the constraints on the Higgs-scalar mixing as the function of the scalar boson mass in Fig.~\ref{FigSun}.
For comparison, the excluded region is also shown from the terrestrial experiments, the meson missing decay~\cite{1809.01876} and the neutron scattering~\cite{2103.05428}.
We adopt the solar interior profile employed in Ref.~\cite{2205.01669}.

For the scalar mass heavier than about 30 keV, the terrestrial
experiments give  the strongest constraint on the mixing angle of
$\sin\th\lesssim 10^{-4}$.
The constraint by the resonant plasma mixing, which is
$\sin\th\lesssim 10^{-9}$ for $m_S\lesssim 0.2$ keV, is about an order of magnitude stronger than the bremsstrahlung constraint.
Furthermore, due to the off-resonant plasma mixing, the constraint is slightly stronger than the $e$-$N$ bremsstrahlung case.
It is found that the absorption effect including the plasma mixing slightly opens the unconstrained region around the large mixing $\sin\th\sim 1$, but this region is already covered by the terrestrial experiments.

Using the standard solar profile, the constraint on the Higgs-scalar mixing from the bremsstrahlung process is reduced by a factor of $1/20$ compared to the one using the constant interior profile.
This profile effect could be roughly reproduced by changing the constant number density shown in Table~\ref{TabStar}.
We find that if the temperature and the density at $r\sim
0.4R_\text{Sun}$ is adopted as the representative values, while the
original ones in Table~\ref{TabStar} are chosen at $r\sim 0.2R_\text{Sun}$, the constant profile gives similar constraints with the standard profile.

\subsection{Horizontal-branch stars}
\label{SecHbs}

\begin{figure}[hbt]
 \centering
 \includegraphics[scale=.3]{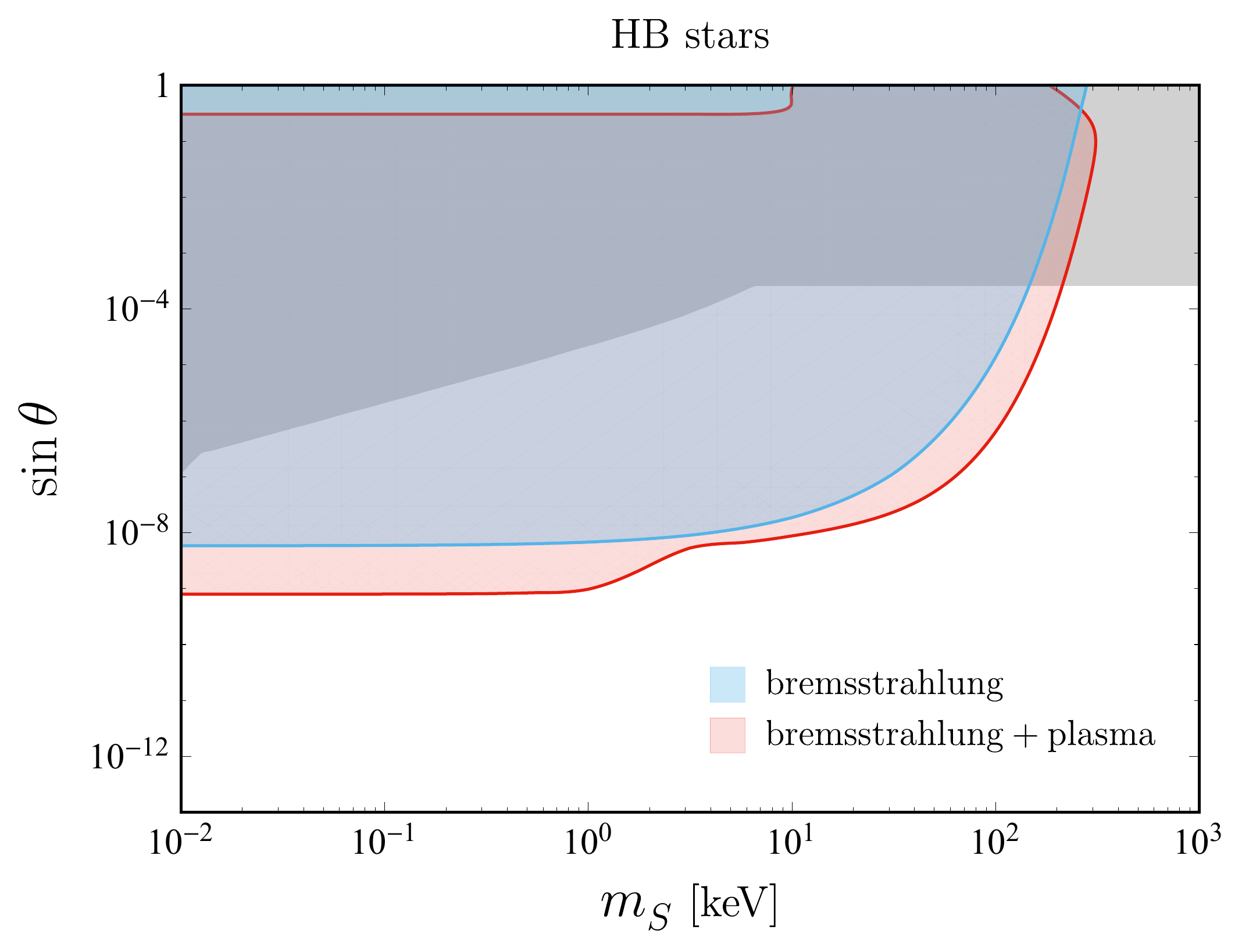}
\caption{
 The limits on the scalar parameters from the horizontal-branch stars.
 The lines are for only the $e$-$N$ bremsstrahlung process (blue) and with the plasma mixing (red).
 The terrestrial experiment bound is the same as in Fig.~\ref{FigSun}.
}\medskip
\label{FigHbs}
\end{figure}

Figure~\ref{FigHbs} shows the limits on the mixing and the mass of the light scalar boson by the cooling argument of horizontal-branch stars.
For the core temperature and density given in Table~\ref{TabStar}, the electron plasma is non-degenerate and non-relativistic.
The results are then similar to the solar constraints.

Since the core temperature and density are higher than those of the Sun, the constraints are extended to the higher mass region of $m_S\sim 200$ keV and the resonant plasma mixing works around $m_S\sim 2$ keV.
For the low-mass region, the bremsstrahlung and the plasma mixing constraints exclude the scalar mixing up to $\sin\th\sim 10^{-8}$ and $10^{-9}$ respectively.
These constraints are similar to those from the Sun including the profile.
The constraint including the plasma mixing effect is roughly consistent with the resonant and the off-resonant limits in Ref.~\cite{1611.05852}.
\subsection{Red giants}
\label{SecRg}

\begin{figure}[hbt]
 \centering
 \includegraphics[scale=.3]{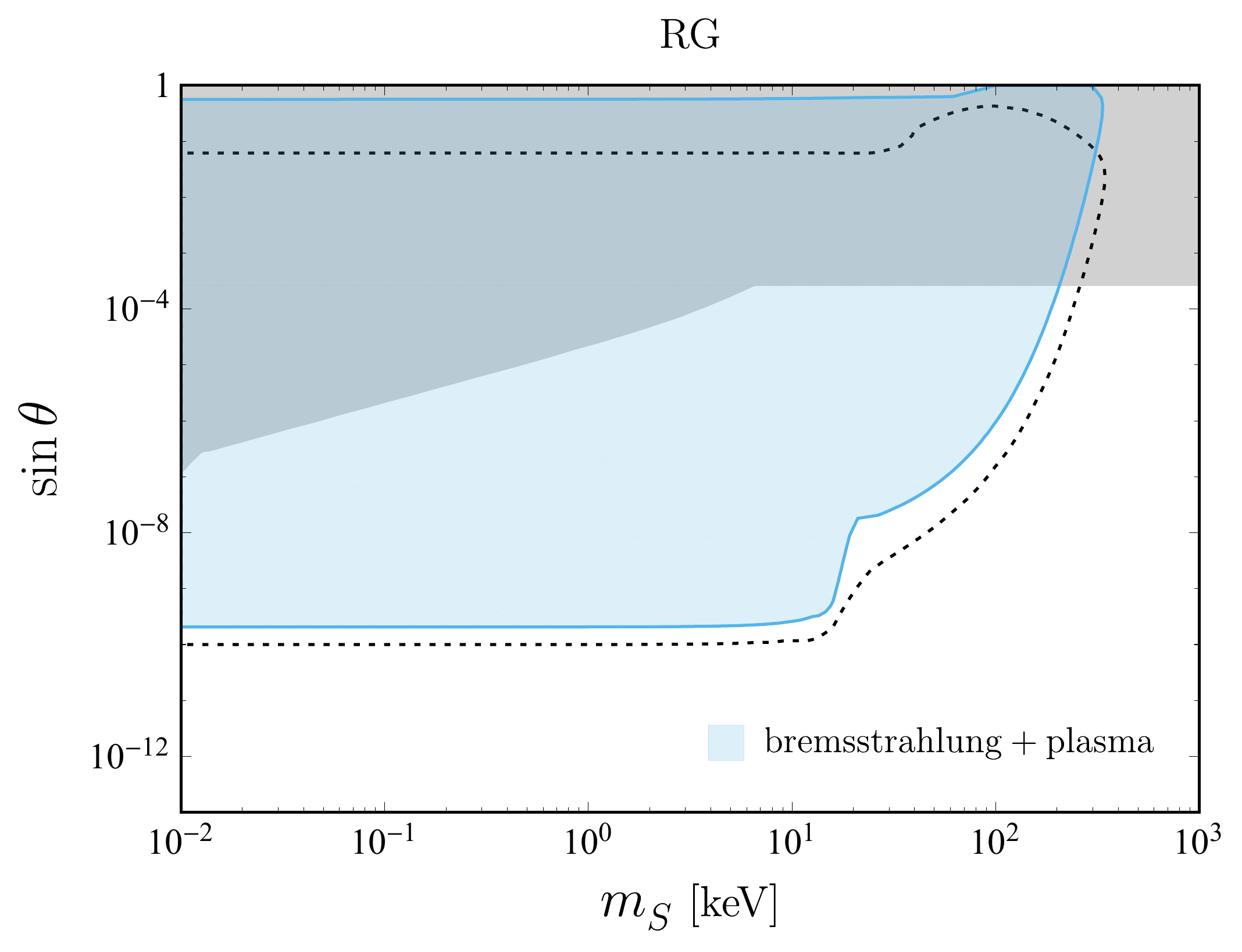}
\caption{
 The scalar parameter limits from the red giants with the bremsstrahlung and the plasma mixing.
 The blue region is excluded by the
 partially degenerate electron plasma resonance and the $e$-$N$ bremsstrahlung
 with the constant profile.
 The dotted line stands for the exclusion by the non-degenerate case,
 which corresponds to the treatment of
 Refs.~\cite{2010.01124,2205.01669} without the plasma mixing.
 The terrestrial experiment bound (gray) is the same as in Fig.~\ref{FigSun}.
}\medskip
\label{FigRg}
\end{figure}

The stellar limit on the Higgs portal mixing angle by red giants is shown in Fig.~\ref{FigRg}.
The density of red giant is high, so electrons in the core are between non-degenerate and degenerate.
We include the resonant plasma mixing from the partially degenerate
electron following Ref.~\cite{1611.05852} without the core profile.
For the bremsstrahlung emission, 
we follow the prescription proposed in Ref.~\cite{hep-ph/9410205} that, 
to evaluate the marginal situation of the red giant core density, 
non-degenerate and degenerate results are interpolated by the inverse sum of their emission rates.
In the current situation, the $e$-$N$ bremsstrahlung bound is equivalent to the following degenerate evaluation.

In the case of the degenerate core, the electron phase space is
suppressed by the Pauli blocking effect.
The degenerate effect is introduced with the approximation where the process is governed by electrons around the Fermi surface, see Ref.~\cite{Raffelt}.
Without including the opacity factor, we find the approximate formula for the energy emission rate from the $e$-$N$ bremsstrahlung process,
\begin{align}
  Q^{eN} \simeq
	\frac{\alpha^2}{3\pi^3} \frac{n_N T^2}{m_N^2}
	(g_N m_e -g_e m_N)^2
  \ln\Big(\frac{\tilde{p}_F^2+k_D^2}{k_D^2}\Big) F(\eta,m_S/T),
\label{EqFsapp}
\end{align}
where the function $F$ is defined as
\begin{align}
  F(p,q) = \int_q^\infty\!\! dx\,
  \frac{(x^2-q^2)^{3/2}}{x^3(e^x-1)}
  \ln\Big(\frac{e^p+1}{e^{p-x}+1}\Big).
\end{align}
The electron degeneracy parameter is given by $\eta=(\mu_e-m_e)/T$, where $\mu_e$ is the electron chemical potential.
For the red giant core, we obtain $\mu_e=6.5\times 10^2$ keV from the temperature and the density given by Table~\ref{TabStar}.
The Fermi momentum is $\tilde{p}_F\sim(3\pi^2n_e)^{1/3}$ and $k_D$ denotes the Debye screening scale which is dominated by heavy ions in the stellar medium.
We consider the screening length $k_D^2= (4\al/\pi)E_F \tilde{p_F}+(4\pi\al/T)\sum_i Z_i^2 n_i$, where $E_F$ is the Fermi energy of electron and $i$ is the nuclide in the plasma~\footnotemark.
\footnotetext{
 This is for the case of weak screening for degenerate electrons while they may be in the strong screening regime at high density.
 Even in this case, however, the weak screening treatment can be a good approximation if the typical distance between nuclear ions $\sim n_N^{-1/3}$ is smaller than the photon wavelength~\cite{Raffelt}.
} 
Including this degeneration effect, the constraint by the bremsstrahlung is about four orders of magnitude weaker than the previous result.
Three of them are due to the correction of the energy emission rate as discussed in Section~\ref{SecBremss}.

The cooling limits become stronger than the terrestrial constraint below about 200 keV.
Since this point depends on the core temperature, it is the same as the constraint imposed by horizontal-branch stars.
The plasma frequency in red giants is high due to the high core density.
Then, the resonant plasma mixing with the partially degenerate electron dominates the bound below 10 keV.
This contribution excludes the scalar mixing close to $\sin\th\sim 10^{-10}$.

For comparison, we show the constraint by the non-degenerate calculation.
This constraint is slightly stronger than that including the partial degeneration.
The non-degenerate result has a gap around the large mixing region, which is opened by the non-resonant mixing.
However, such a gap region is already excluded by the terrestrial experiments.
%
\subsection{White dwarfs}
\label{SecWd}

\begin{figure}[hbt]
 \centering
 \includegraphics[scale=.3]{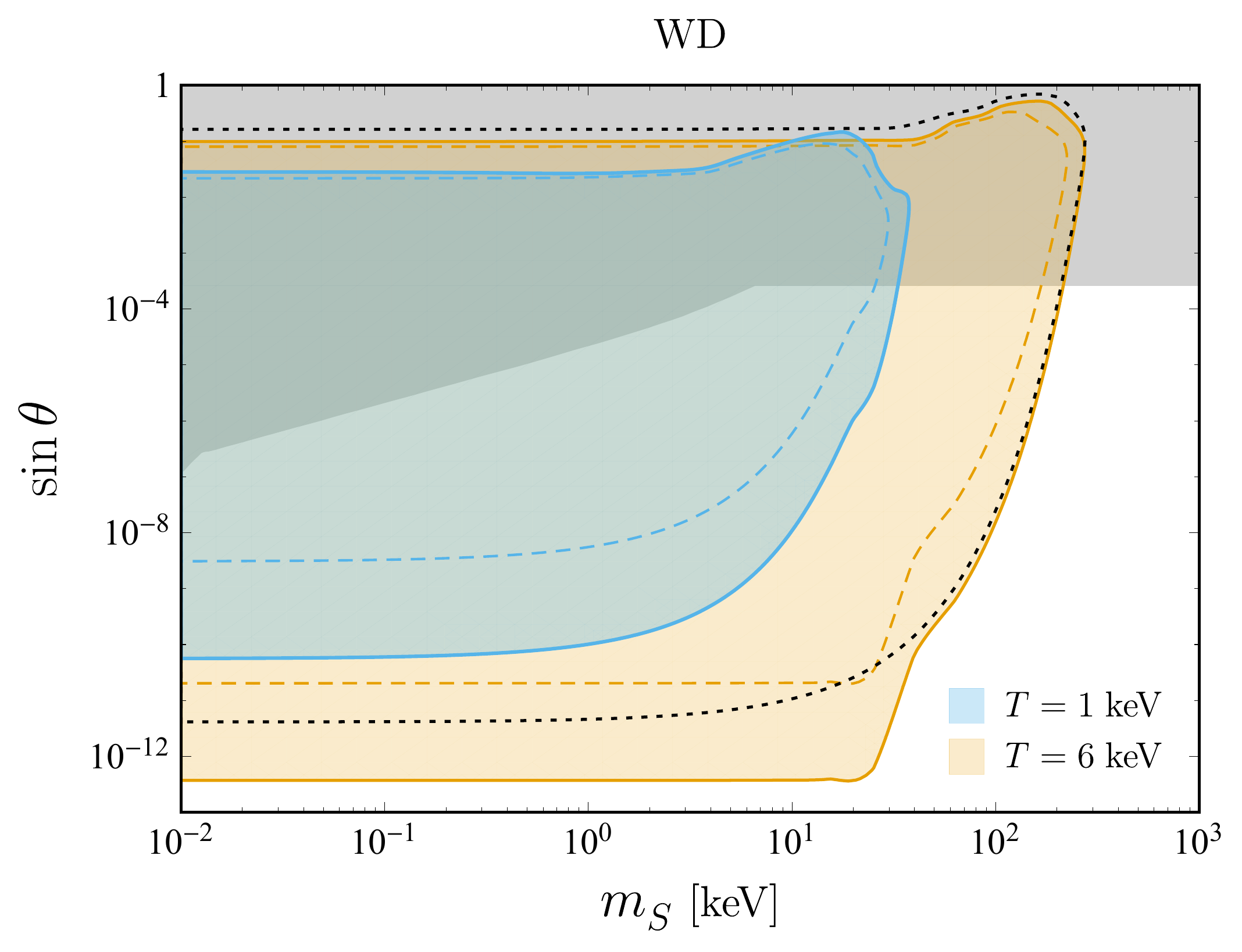}
\caption{
 The limits on the scalar parameters from white dwarfs.
 For the core temperature, we adopt two benchmark values $T=1$~keV (blue) and $T=6$~keV (orange).
 The solid and dashed lines indicate two different upper limits for the luminosity, $10^{-5}\,L_\text{Sun}$ and $0.03\,L_\text{Sun}$, respectively.
 The black dotted line stands for the exclusion boundary from the
 luminosity limit of $10^{-5}\,L_\text{Sun}$ only with the
 bremsstrahlung.
 The terrestrial experiment bound (gray) is the same as in Fig.~\ref{FigSun}.
}\medskip
\label{FigWd}
\end{figure}

The cooling limits from white dwarfs are summarized in Fig.~\ref{FigWd}.
We obtain the bounds with the $e$-$N$ bremsstrahlung and the plasma mixing in the highly degenerate electron plasma.
In the following calculation, the degeneration effect is introduced
with the Fermi surface approximation, as needed, with $\mu_e=7.9\times 10^2$ keV which can be computed with the temperature and density given in Table~\ref{TabStar}.

In Refs.~\cite{2010.01124,2205.01669}, the constraints on the scalar
parameters are evaluated with the core temperature of 6 keV based on Ref.~\cite{Ref6kev}.
In addition to this temperature, we also show the constraints with the core temperature of 1 keV following Refs.~\cite{Raffelt}.
We also evaluate two constraints corresponding to the luminosity upper limits of $10^{-5} L_\text{Sun}$ and $0.03 L_\text{Sun}$.
The 6 keV constraint from the bremsstrahlung with the upper limit of $10^{-5} L_\text{Sun}$, which is shown by the black dotted line in Fig.~\ref{FigWd}, is about six orders of magnitude weaker than the previous results.
About four of them can be explained by the correction of the energy emission rate because of the additional enhancement by the nuclear charges.

The constraints with $T=6$ keV and the 1 keV become stronger than the terrestrial constraint below $m_S\sim 200$ keV and 30 keV, respectively.
The lower the core temperature becomes, the weaker the parameter bounds.
The resonant mixing effect can be neglected for $T=1$ keV in the degenerate case.
Then, the $T=1$ keV constraints are about two orders of magnitude weaker than the 6 keV constraints.
The former constraint excludes the Higgs-scalar mixing angle up to $\sin\th\sim 10^{-10}$ for the limit of $L<10^{-5}L_\text{Sun}$ and $\sin\th\sim 10^{-8.5}$ for $L<0.03 L_\text{Sun}$ in the region of $m_S\lesssim$1 keV\@.
The latter constraint with $T=6$ keV excludes the mixing larger than
about $10^{-12.5}$ and $10^{-11}$, respectively, in the same low mass region.

\section{Conclusion}
\label{SecCon}

We have re-evaluated the $e$-$N$ bremsstrahlung contributions to the stellar cooling limits on the Higgs portal light scalar boson.
We also include the plasma mixing effect, which often gives comparable or even stronger limits than the bremsstrahlung. 
Compared to the previous study, the scalar emission rate from the $e$-$N$ bremsstrahlung process is found to be suppressed by the square of the electron-nucleus mass ratio, and then the parameter limits become weaker. 
Furthermore, in the case of dense cores where electrons are degenerate, the emission rates are generally reduced by the Pauli
blocking effect, that also makes the limits weaker. 
\begin{figure}[hbt]
 \centering
 \includegraphics[scale=.4]{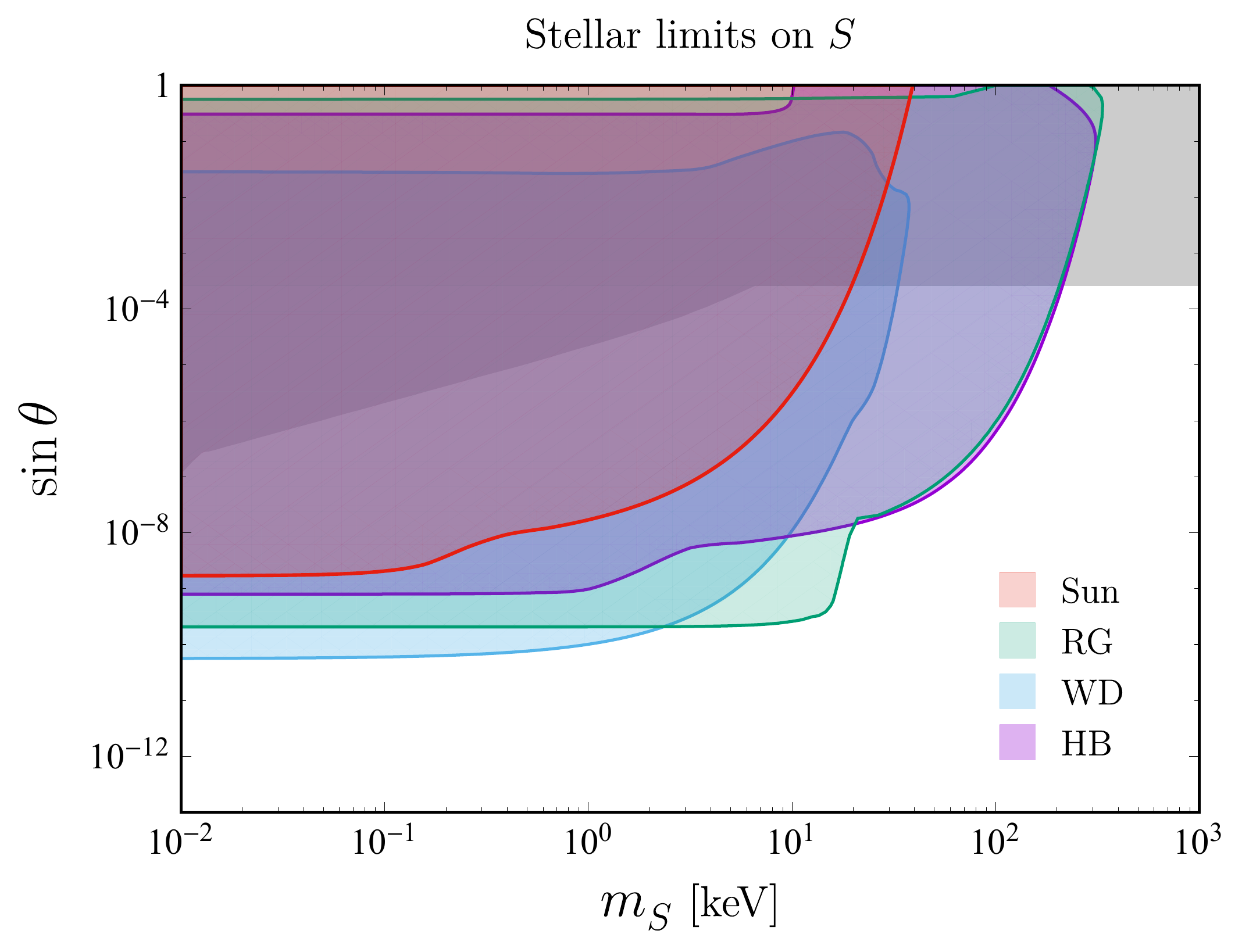}
 \caption{
  Summary of the stellar limits on the Higgs-portal scalar parameters: the mass $m_S$ and the mixing angle $\theta$.
	The colored regions are excluded by the stellar cooling arguments for the Sun, red giants (RG), white dwarfs (WD), and horizontal-branch (HB) stars.
	For the details of lines, see the text or the corresponding sections.
 The terrestrial experiment bound (gray) is the same as in Fig.~\ref{FigSun}.
 }\medskip
\label{FigSum}
\end{figure}

The summery of the stellar cooling limits in the mixing-mass plane for the light scalar is shown in Fig.~\ref{FigSum}.
For the Sun and horizontal-branch stars, we show
the constraint including the plasma mixing effect and the solar profile.
For red giants, the bound is obtained mainly from the partially degenerate
resonant plasma mixing in wider region. 
For white dwarfs, we show the bound with the core
temperature $T=1$ keV, for which the bremsstrahlung process leads to
the bound. For all the lines in the figure, 
the lower boundaries in low-mass region are understood to be
proportional to the square roots of the luminosity upper limits.

For the heavier scalar mass region, the high-temperature stars such as
horizontal-branch stars and red giants put stronger bounds than the terrestrial experiments up to $m_S\sim 200$ keV.
For the lower mass region, the stellar cooling arguments are found to 
exclude the mixing angle up to $10^{-9}$--$10^{-10}$. 
For all the constraints, the regions where the absorption effect in the medium works well are completely excluded by the terrestrial bound.

The obtained exclusion regions could vary by an order of magnitude, depending on the stellar conditions and the adopted assumptions. 
To reduce the uncertainties in the constraints, we would need to include more details of the stars and processes in future calculation.
\\[1em]
\textbf{Note added}: During the final stages of preparing this manuscript, the paper \cite{2303.00778} appeared on arXiv.
This paper claims the $(m_e/m_N)^2$ suppression of the scalar emission by nuclei, which is the same as our present result, compared to Refs.~\cite{2010.01124,2205.01669}.
%
\section*{Acknowledgments}
The work of Y.Y. was supported by the National Science and Technology Council, the Ministry of Education (Higher Education Sprout Project NTU-112L104022), the National Center for Theoretical Sciences of Taiwan, and the visitor program of Yukawa Institute for Theoretical Physics, Kyoto University.
The work of K.Y. was supported in part by Japan Society for the Promotion of Science KAKENHI Grant Number JP20K03949.

\appendix
\section{Energy emission rates from bremsstrahlung}
\label{AppBremss}

We here present the formulas for the energy emission rates per unit time and volume from the scalar bremsstrahlung processes such as $eN\to eNS$, $ee\to eeS$ and $NN\to NNS$. 
Their formal expression is given by \eqref{Qgeneral} and \eqref{Rgeneral}.
The following emission rates are evaluated in the approximation that the electrons and nuclei in the stellar medium are non-relativistic. 
We then find the energy emission rate of the $A$-$B$ bremsstrahlung process by partially performing the phase space integral,
\begin{align}
  Q^{AB} &= 
  \frac{n_An_B\,M^{3/2}m^3\,T^{1/2}}{2^5(2\pi)^{9/2} (m_Am_B)^{7/2}} 
  \int^\infty_q \! du\,  \int^\infty_0 \! dv\, \int^1_{-1}\!dz
   \int \! d^3k_S \;\sum|\mathcal{M}|^2 
 \sqrt{uv}\,e^{-u} \delta(u-v-x) ,
\end{align}
where $M$ and $m$ are the total and reduced masses defined by the masses of the external particles as $M=m_A+m_B$ and $m=m_Am_B/(m_A+m_B)$.
The integral variables are $u=p_I^2/(2mT)$, $v=p_F^2/(2mT)$, and $z$ is the angle between $\vec{p}_I$ and $\vec{p}_F$, where the ``relative'' momenta $p_{I,F}$ are defined by $Mp_I=m_Bp_A-m_Ap_B$ and $Mp_F=m_Bp_A'-m_Ap_B'$.
By multiplying the opacity factor and integrating the scalar momentum (energy) with the corresponding matrix element $\mathcal{M}$, the explicit form is derived. 

For the scalar coupled to different initial states $A$ and $B$ with the couplings $g_A$ and $g_B$ respectively, the energy emission by the $A$-$B$ bremsstrahlung process is
\begin{align}
 Q^{AB} = \frac{2^3 \al^2 n_A n_B M^{3/2}m^2
 T^{1/2}}{3(2\pi)^{3/2} (m_Am_B)^{7/2}}
 (g_A m_B -g_B m_A)^2 \, I(m_S/T).
\end{align}
The function $I$ is defined in \eqref{EqIntbremss}. 
Applying it to the scalar which couples to the electron and the nucleon (then $M\simeq m_N$ and $m\simeq m_e$), we obtain the explicit form of Eq.~\eqref{EqBremss} without including the opacity factor. 
For a heavier nucleus $N_i$, the rate for the $e$-$N_i$ bremsstrahlung process is obtained by the replacements $m_N\to A_im_N$, $\,g_N\to A_ig_N$, $\,n_N\to n_{N_i}$, and $\alpha\to Z_i\alpha$ as noted in the text. 
In this case, the dependence of the mass number $A_i$ is dropped from the above emission rate leaving the $Z_i^2$ factor. 

Similarly, we find the emission rates for the $e$-$e$ and $N$-$N$ bremsstrahlung processes mediated by the photon,
\begin{align}
  Q^{ee} &= 
  \frac{2\al^2 n_e^2\,T^{3/2}}{15\pi^{3/2} m_e^{5/2}} \,
  g_e^2 \,J(m_S/T) ,  \\
  Q^{NN} &= 
  \frac{2\al^2 n_N^2\,T^{3/2}}{15\pi^{3/2} m_N^{5/2}} \,
  g_N^2 \,J(m_S/T) .
\end{align}
The function $J$ is defined as,
\begin{align}
  J(q) &= \int^\infty_q\! du \int^\infty_0\! dv \int^1_{-1}\! dz
  \int^\infty_q\! dx \; e^{-u}\sqrt{uv}\,\delta(u-v-x)
  \nonumber \\[1mm]
  &\hspace*{10mm}  
  \frac{(u+v)^2+12uvz^2}{((u+v)^2-4uvz^2)^2}
  \Big(1-\frac{q^2}{x^2}\Big)^{3/2} \Big[ 
  3\big[(u+v)^2-4uvz^2\big]\Big(1-\frac{q^2}{x^2}\Big) +4x^2+q^2 \Big].
\end{align}
In the $e$-$e$ and $N$-$N$ cases, since the two external states are the same species of particles, the leading contribution to the squared matrix elements is canceled out.
Then, the emission rates are more suppressed by the mass scales. 
For a heavy nucleus $N_i$, the $S$ energy emission rate from the $N_i$-$N_i$ bremsstrahlung process is obtained by the replacements $m_N\to A_im_N$, $\,g_N\to A_ig_N$, $\,n_N\to n_{N_i}$, and $\alpha\to Z_i^2\alpha$ in the above $Q^{NN}$ formula.
This replacement leads to the suppression by $1/\sqrt{A_i}$ and the enhancement by the atomic number as $Z_i^4$, compared to the $N$-$N$ case.
In Fig.~\ref{FigEmisun}, we show the comparison of different emission rates in the solar environment.
\bigskip


\end{document}